\documentclass[aps,prb,showpacs,amsmath,amssymb,twocolumn]{revtex4}
\pdfoutput=1

\usepackage{xcolor}
\usepackage{graphicx}
\usepackage{amsmath}
\usepackage{amssymb}
\usepackage{amsfonts}
\usepackage{units}
\usepackage{bm}
\usepackage{hyperref}
\usepackage{enumerate}

\usepackage{longtable}
\usepackage{totcount}

\usepackage{xcolor}

\definecolor{MS-color}{RGB}{255,0,0}

\begin{document}

\title{Multiple optical gaps and laser with magnonic pumping in 2D Ising superconductors }

\date{\today}

\author{Mikhail Silaev}
\affiliation{Computational Physics Laboratory, Tampere University, Tampere 33720, Finland}
\affiliation{Moscow Institute of Physics and Technology, Dolgoprudny, 141700 Russia}
\affiliation{Institute for Physics of Microstructures, Russian Academy of Sciences, 603950 Nizhny Novgorod, GSP-105, Russia}

\begin{abstract}
Ising superconductivity has been  recently discovered in 2D transition metal dichalcogenides.
We report that such superconductors have unusual optical properties  controlled by the in-plane Zeeman field. 
First, we find several optical gaps  visible as peaks of the  conductivity and the Raman susceptibility. 
Moreover, we find that the Ising spin splitting in the spectrum of Bogolubov quasiparticles
enables  strong population inversion generated by the time-dependent Zeeman field.  
Ultimately this leads to the possibility of the superconducting laser with magnonic pumping 
which can be realized in the van der Waals structures consisting of the Ising superconductor and the ferromagnetic insulator layers.
 \end{abstract}
 
 \maketitle
 It is difficult to overestimate the importance of the optical probes for studies of superconducting materials. The existence of superconducting gap in the quasiparticle spectrum has been discovered with the  help of the far-infrared  optical conductivity\cite{glover1957conductivity,mattis1958theory,bardeen1957theory}. 
 Later on the superconducting gap has been observed by the inelastic Raman scattering
 \cite{sooryakumar1980raman,
hackl1983gap,dierker1983electronic},
 which has become one of the basic tools in studying the anisotropic 
 \cite{devereaux1994electronic,
devereaux1995electronic} and multiband superconductivity\cite{
chen2001evidence,
chubukov2009theory,udina2020raman} as well as the superconducting collective modes\cite{sooryakumar1980raman,
littlewood1981gauge,
littlewood1982amplitude,
blumberg2007observation,
cea2016signature,
grasset2018higgs,
measson2014amplitude,
leroux2012anharmonic,
grasset2019pressure}. 
   
Previous  studies of the optical properties have been focused on the 3D materials. Recently the  family of 2D superconducting materials has appeared\cite{saito2016highly},  in twisted bilayer graphene\cite{cao2018unconventional,
lu2019superconductors,
yankowitz2019tuning}, rhombohedral trilayer graphene\cite{Zhou2021} and few-layer transition metal dichalcogenides (TMD) \cite{xi2016ising, lu2015evidence, tsen2016nature,ugeda2016characterization, staley2009electric, saito2016superconductivity, lu2018full}. The in-plane symmetry breaking combined with the  heavy transition-metal atoms leads to the extremely large spin-orbital coupling in TMD \cite{mak2010atomically,
radisavljevic2011single, wang2012electronics,
zhu2011giant,xiao2012coupled} (SOC). It  has the Ising character having the form of the effective out-of-plane Zeeman field changing the sign between electron pockets at $K$ and $K^\prime$ points of the Brillouin zone\cite{
zhu2011giant, xiao2012coupled, lu2015evidence, zhou2016ising, xi2016ising, wang2012electronics}. 
%
   %
  Recently discovered so-called Ising superconductivity (IS) \cite{lu2015evidence, zhou2016ising,xi2016ising,xing2017ising, li2021printable} 
  features  strong enhancement of the in-plane critical field\cite{sohn2018unusual,lu2015evidence, zhou2016ising,sergio2018tuning,kuzmanovic2021tunneling,saito2016superconductivity}, 
 giant thickness-dependent transition state spin Hall signal\cite{jeon2021role},  electric field effect on superconductivity\cite{costanzo2018tunnelling, lu2015evidence},
  unconventional supercurrent phase\cite{idzuchi2021unconventional,idzuchi2020van}, multiple gaps\cite{dvir2018spectroscopy}, collective modes\cite{wan2021observation},   and  
  non-trivial interplay with magnetism \cite{hamill2021two,idzuchi2020van, ai2021van, kezilebieke2020moir,kezilebieke2020topological,kang2021giant, kim2019tailored,cho2020distinct}. 
  %
 Theoretically such superconductors have been studied in the context of the unconventional pairing mechanisms
 \cite{wickramaratne2021ising, ilic2017enhancement, mockli2020ising, mockli2018robust, mockli2020magnetic, haim2020signatures, 
mockli2019magnetic, liu2020microscopic},
 magnetic properties\cite{ilic2017enhancement, mockli2020ising, tang2021magnetic}, parity\cite{mockli2019magnetic,mockli2018robust,mockli2020ising,rahimi2017unconventional,wickramaratne2020ising} and singlet-triplet mixing superconductivity \cite{haim2020signatures,mockli2018robust,mockli2019magnetic,mockli2020ising,wickramaratne2020ising,rahimi2017unconventional}, topological \cite{zhou2016ising}, transport properties\cite{cheng2019switch,tang2021controlling} and magnetic properties \cite{wickramaratne2020ising,aikebaier2021controlling,wickramaratne2021magnetism}.  
 
  
 
 \begin{figure}[htb!]
 \centerline{$
 \begin{array}{c} 
 \includegraphics[width=1.0\linewidth]
 {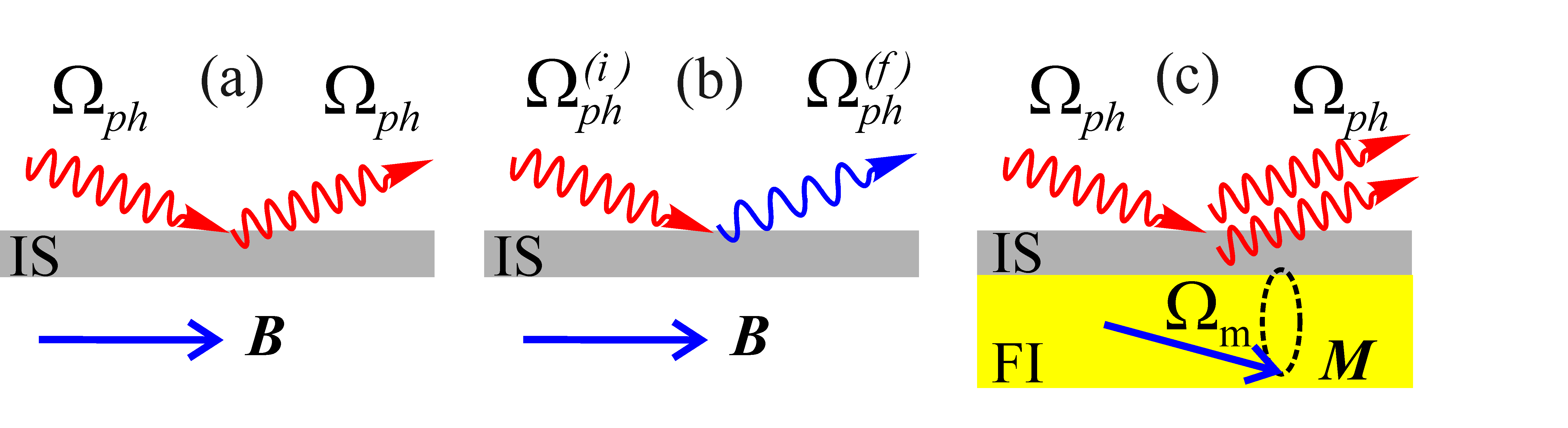} 
  \end{array}$}
 \caption{\label{Fig:1} 
 Optical properties of the IS. (a) Photon reflection with the frequency $\Omega_{ph}$.
 (b) Raman scattering from the incident to the reflected photon of the frequencies $\Omega^{(i)}_{ph}$ and $\Omega^{(f)}_{ph}$, respectively. 
 (c) Superconducting laser with magnon pumping generated by the ferromagnetic insulator (FI) with the magnetization $\bm M$ precessing at frequency
 $\Omega_m$.  }
 \end{figure}   
 
 One can expect that 2D IS should have  unusual optical properties but there have been no studies in this direction.
In the present Letter we predict 
   multiple peaks in the spectrum of the  optical conductivity $\sigma (\Omega_{ph})$ and the Raman scattering $\chi_{Ram} (\Omega_{Ram})$ where $\Omega_{Ram} =  \Omega^{(f)}_{ph} - \Omega^{(i)}_{ph} $ is the Raman shift, see Fig.\ref{Fig:1}a,b.  
  These features are controlled by the on-plane Zeeman field which can be induced either by the external magnetic field or by the magnetic proximity effect in the Van der Waals (VdW) structure consisting of the superconductor and ferromagnetic insulator (FI) layers. 
  Recently the VdW  IS/FI structures like NbSe$_2$/ CrB$_3$ or other combinations has been studied experimentally
  \cite{hamill2021two,idzuchi2020van, ai2021van, kezilebieke2020moir,
  kezilebieke2020topological,kang2021giant, kim2019tailored,cho2020distinct}
  and  theoretically\cite{wickramaratne2021magnetism}. 
    The optical response of IS pumped by magnons Fig.\ref{Fig:1}c, that is by the time-dependent Zeeman field features the lasing effect. 
  Ising spin splitting of Bogolubov spectrum allows to generate very strong  quasiparticle population inversion resulting  in the negative conductivity $Re (\sigma) (\Omega_{ph})<0$ of the IS. 
  
 \begin{figure}[htb!]
 \centerline{$
 \begin{array}{c} 
 \includegraphics[width=1.0\linewidth]
 {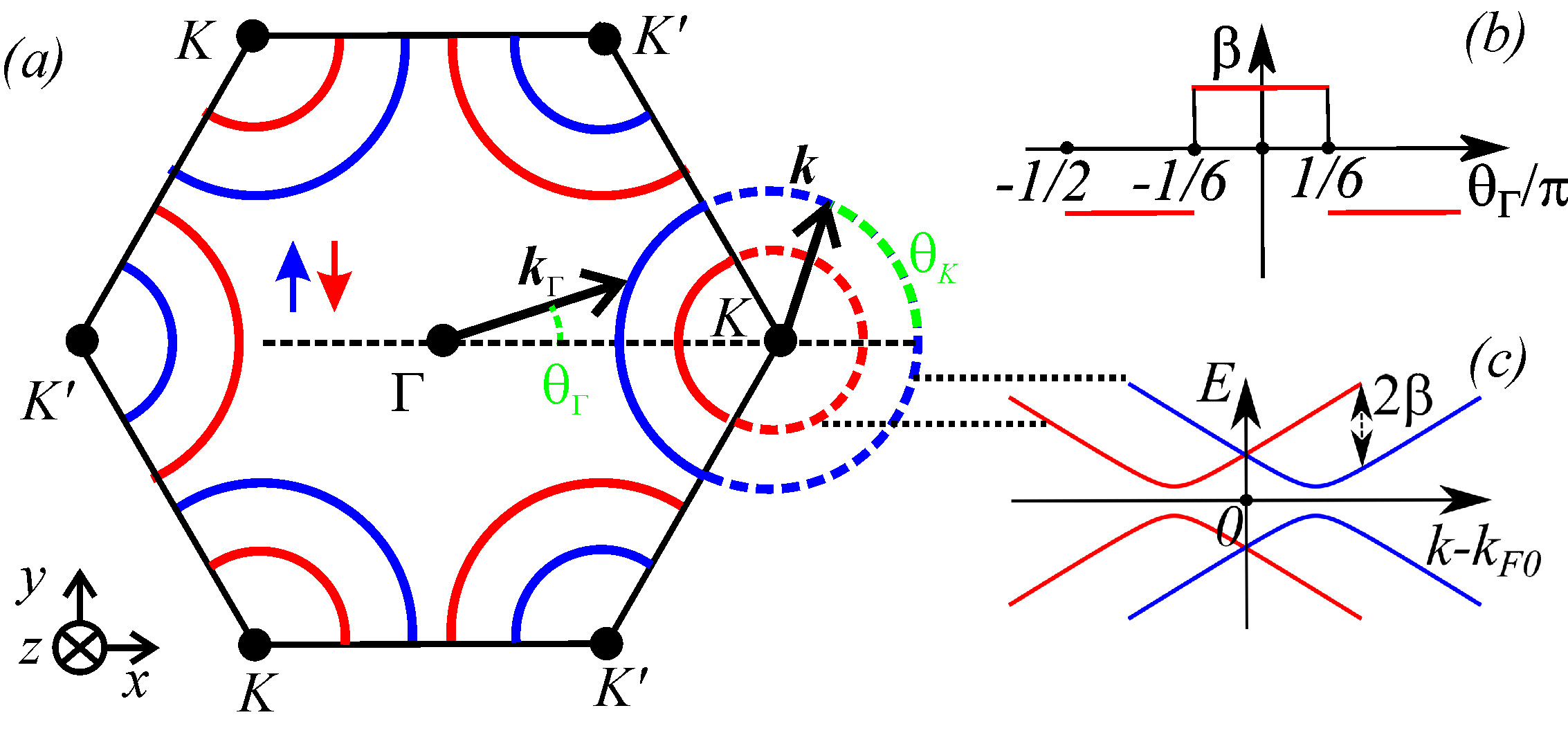} 
  \end{array}$}
 \caption{\label{Fig:Model} (a) Schematic Fermi surface of the two-dimensional metal with Ising SOC field oriented along $z$ causing the spin-splitting between spin-up (blue) and spin-down (red) states. The parts of Fermi surface in the second Brillouin zone are shown by  dashed lines. 
 (b) The model piecewise-constant  dependence of the SOC strength  $\beta(\theta_\Gamma)$.  
 (c) Bogolubov quasiparticle spectrum in the Fermi surface pocket near the $K$ point. The red and blue lines correspond to the spin-up and spin-down Fermi surfaces as in (a).      
  }
 \end{figure}   
  
 Our model consists of the multiband 2D superconductor with the pairing between the electronic states situated at $K$ and $K^\prime$ points of the Brillouin zone as shown in Fig.\ref{Fig:Model}a, with the Zeeman field is $\bm h$.
  It is described by the Hamiltonian
  \begin{align} \label{Eq:BdGHam}
 & \hat H =\hat \tau_3 
 (\xi_k - i \hat \Lambda )  
  \\
  & \hat\Lambda = i (\beta \hat\tau_0 \hat\sigma_z + 
   \hat\tau_3 \bm h \bm {\hat\sigma} ) + \hat\tau_1\Delta  .
  \end{align}
  where $\hat\tau_{1,2,3}$ and $\hat\sigma_{x,y,z}$ are the Pauli matrices in Nambu and spin spaces, respectively.
  The kinetic energy relative to the Fermi level is $\xi_k = (k^2 -k^2_{F0})/2m = \bm v_F \bm k  $, where the Fermi velocity $\bm v_F$ in general depends on the momentum direction. The SOC field which in general has to be the odd function of momentum direction relative to the $\Gamma$ point $\beta(\bm k_\Gamma)= - \beta(-\bm k_\Gamma)$ is taken in the model piece-wise constant form shown in Fig.\ref{Fig:Model}b.  %
  The order parameter $\Delta$ is assumed to the isotropic.
  The in-plane Zeeman field $h \bm x$ is induced either by the external magnetic field $h = \mu_B B$ or by the magnetic proximity effect with the adjacent FI.  
  
\begin{figure}[htb!]
 \centerline{$
 \begin{array}{c} 
  \includegraphics[width=0.5\linewidth]
 {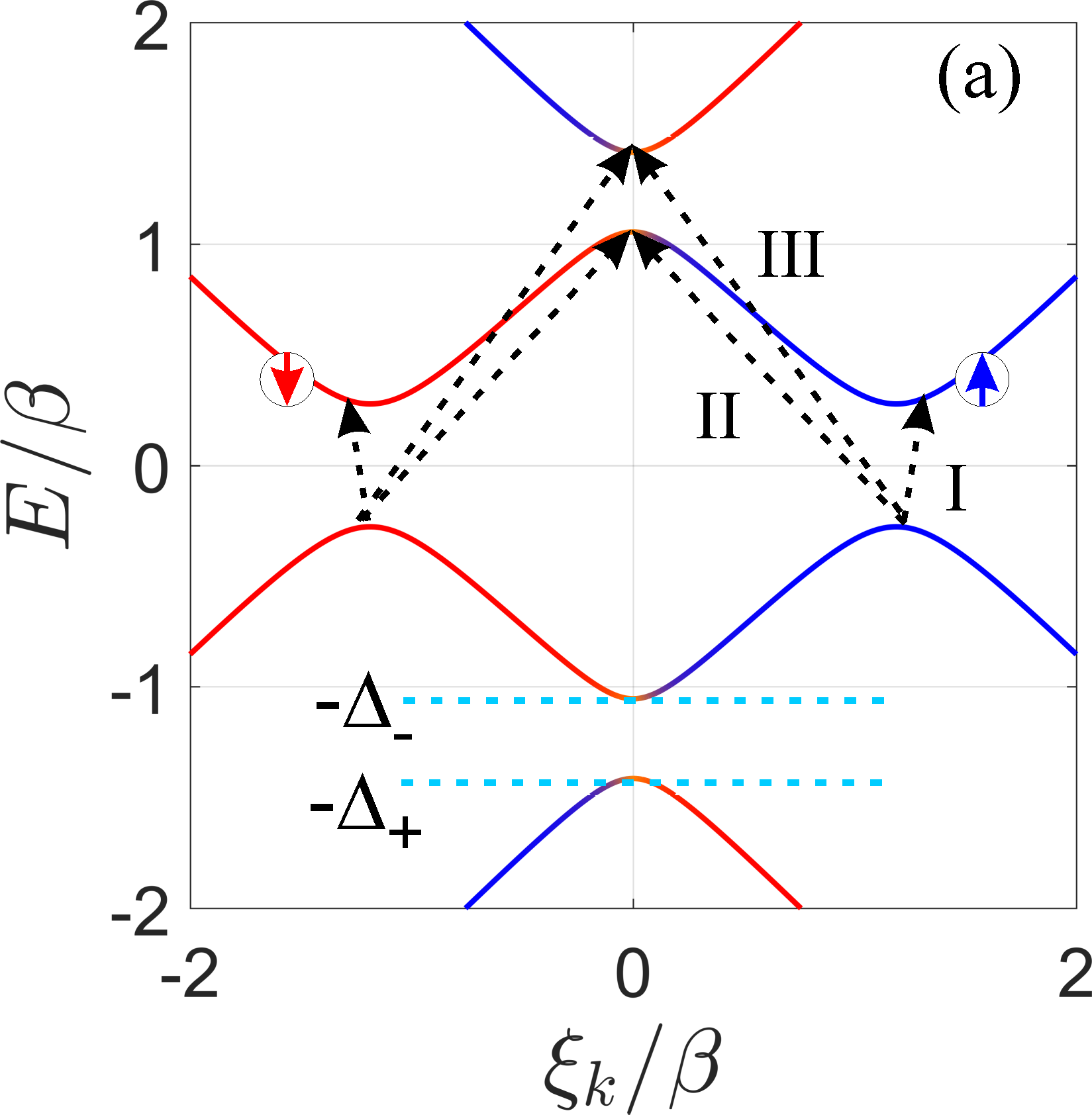}
 \includegraphics[width=0.5\linewidth]
 {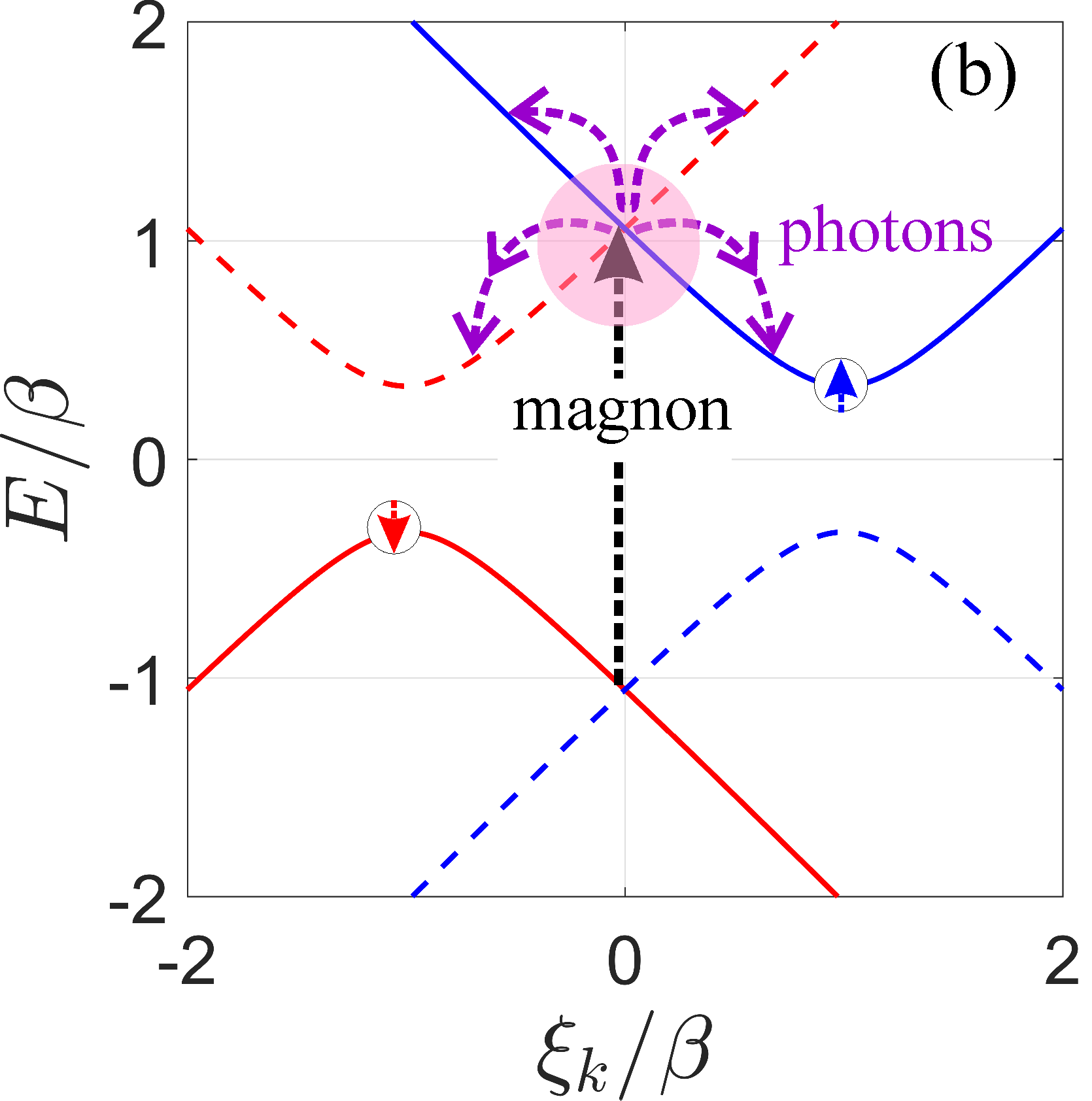}
 \end{array}$}
 \caption{\label{Fig:Spectrum} (a) 
 Bogolubov spectrum of the IS with in-plane Zeeman field $h \bm x \neq 0$. The color code of the branches is determined by the average spin projection $\langle S_z\rangle$ projection ranging from $+1/2$ to $-1/2$ shown by the blue and red colors respectively. The dashed arrows $I$, $II$, $III$ show excitation processes corresponding to the  optical gaps $\Omega_I$, $\Omega_{II}$, $\Omega_{III}$.  
 (b) Lasing effect in the IS with magnon pumping. The vertical black arrow shows the excitation of quasiparticles by the oscillating Zeeman field $h_m \cos(\Omega_m t) \bm x$ generating hot spot in the quasiparticle population (pink shading). The violet  arrows  show  photon emissions or absorptions  with quasiparticle down/up transitions assisted by the impurity scattering. }
 \end{figure}

  The Hamiltonian (\ref {Eq:BdGHam} ) yields Bogolubov spectrum consisting of two spin-split branches 
  \begin{align}\label{Eq:BogolubovSpecrum}
    E_{\uparrow (\downarrow)} (k) =  \sqrt{E_{bdg}^2 + \beta^2+ h^2
   \pm 2\sqrt{h^2E_{bdg}^2 + \xi_k^2 \beta^2 }}
  \end{align}
  $E_{bdg} = \sqrt{\xi_k^2 + \Delta^2}$ is the  Bogolubov spectrum in the usual superconductor. 
  { For $h=0$ the Ising SOC splits the usual Bogolubov spectrum by shifting the spin-up/down branches left/right as shown in Fig.\ref{Fig:Model}c. 
  }
  The spin-up/down  branch intersection at $k=k_{F0}$ shown in Fig.\ref{Fig:Model}c, that is at $\xi_k=0$ is removed by the Zeeman field $\bm h \perp \bm z$.
 For $h\neq 0$ the spectrum (\ref{Eq:BogolubovSpecrum}) has two types of gaps. The ordinary gap has the magnitude of $\Delta_{ord} = \Delta \sqrt{1-h^2/\beta^2}$ and is located at $\xi_k = \sqrt{\beta^2 - h^2\Delta^2}/\beta$. 
 Note that the spectrum becomes gapless $\Delta_{ord}=0$ at $h>\beta$ while the order parameter remains non-zero $\Delta\neq 0$.  
 The additional so-called "mirage" gap\cite{tang2021magnetic} is located at $\xi_p=0$ and separates the gap edges at $\Delta_\pm =\sqrt{(h\pm \Delta)^2 + \beta^2}$.  
 The spectrum in the illustrative case $\beta=3\Delta$, $h=2\Delta$ is shown in Fig.\ref{Fig:Spectrum}a.
 
 The average $\langle S_z\rangle$ spin projection is shown by the color code of the spectrum branches in Fig.\ref{Fig:Spectrum}a. Near the usual gap edge the spin state is almost $\langle S_z\rangle=\pm 1/2$. On the other hand, near the mirage gap edges $\xi_k=0$ the spin state $\langle S_z\rangle=0$, which indicates the dominating role of the spin-triplet pairing at this point. 
    
Several van Hove singularities in the spectrum (\ref{Eq:BogolubovSpecrum}) results in the multiple 
{\it optical gaps}
visible as peaks in the optical responses. In Fig.\ref{Fig:Spectrum}a three  most important of them are indicated by the dashed arrows  $I$, $II$ and $III$. 
 The excitation process $I$ corresponds to the  breaking of the usual spin-singlet Cooper pairs, while $II$ and $III$ corresponds to the breaking of the spin-triplet Cooper pairs. The corresponding optical gaps are  $\Omega_I = 2\Delta_{ord}$, $\Omega_{II} = \Delta_{ord} + \Delta_-$ and $\Omega_{II} = \Delta_{ord} + \Delta_+ $ respectively. At such frequencies the external perturbation creates the pair of quasiparticles at the van Hove singularities which will be shown to result in peaks in the optical responses.  
 
 For $h=0$ in Eq.(\ref{Eq:BogolubovSpecrum}) the Ising spin-splitting makes the minimal energy difference between empty spin-up and occupied spin-down branches to be $min ( E_\uparrow + E_\downarrow) = 2\beta$.
 Therefore, magnon pumping  with frequency $\Omega_m\approx  2\beta$ creates the hot spot in the quasiparticle population inversion near $\xi_k=0$.  
 This leads to the possibility of the quasiparticle transitions between  $\xi_k=0$ and $\xi_k\neq 0$ states assisted by the impurity scattering and accompanied either by the photon radiation or absorption. Since the density of states increases  at lower energies $N(\varepsilon<\beta)>N(\varepsilon>\beta)$ 
  there are more emission processes resulting in the lasing which shows up as the negative conductivity ${\rm Re} \sigma < 0$. 
 
  %
    
 In each of the six Fermi pockets  in Fig.\ref{Fig:Model}a we  introduce the  quasiclassical propagator \cite{supplement} 
$\hat g = \hat g (t,t^\prime,\bm r,\bm v_F)$
     which satisfies by the Eilenberger equation \cite{Eilenberger1968}
    \begin{align} \label{Eq:Eilenberger}
    & i\bm v_F [\hat\tau_3\bm A,\hat g]_t + 
    [\bm A\hat \gamma \bm A,\hat g]_t    = 
    \\ \nonumber
    & - \{ \hat \tau_3\partial_t, \hat g \}_t + 
    [\hat \Lambda, \hat g]_t + 
    \frac{[\langle \hat g\rangle \circ, \hat g ]_t}{2\tau_{imp}} .
    \end{align}  
        In general it contains the time-dependent vector  potential $\bm A (t)$ and  the Zeeman field $\bm h=\bm h(t)$.  %
    The last term in Eq.\ref{Eq:Eilenberger} is the collision integral describing the intraband impurity scattering with the rate $\tau_{imp}^{-1}$. 
    We denote the commutators 
    $[X,g]_t= X(t) g(t,t^\prime) - g(t,t^\prime)X(t^\prime)$ 
    and the convolution is given by 
    $\langle \hat g\rangle\circ \hat g  = 
    \int d t_1 \langle \hat g\rangle(t,t_1) \hat g (t_1,t^\prime) $. 
    The angle-averaging is done over the  Fermi pocket centered at the $K$ point at $\theta_\Gamma=0$ shown in Fig.\ref{Fig:Model}a  $\langle \hat g\rangle = \int_0^{2\pi} \hat g (\theta_K) d\theta_K $, where $\theta_K$ is the angle which characterises the direction of the momentum $\bm k $ relative to the $K$ point. Here we take into account the contribution of all three $K$ points in Fig.\ref{Fig:Model}a by extending the integral over the angle $\theta_K$ from $0$ to $2\pi$.  
    
The first term in the l.h.s. of Eq.\ref{Eq:Eilenberger} is the usual paramagnetic vartex which determines the conductivity and the Meissner effect.
The second term is the diamagnetic vertex describing  the electron density modulation by the time-dependent $\bm A (t)$. 
This term comes with the coefficient \cite{abrikosov1973theory} $\gamma_{ij} = \partial E/\partial k_i\partial k_i  - \langle \partial E/\partial k_i\partial k_i \rangle$  where the subtraction of Fermi pocket average term takes into account Coulomb interaction. 
The charge neutrality in Eq.\ref{Eq:Eilenberger} is maintained since  $\langle \hat \gamma\rangle =0$. However the direction-dependent density variation responsible for the Raman scattering\cite{klein1984theory,falkovsky1993inelastic}is non-zero.  

 
 \begin{figure}[htb!]
 \centerline{$
 \begin{array}{c} 
  \includegraphics[width=1.0\linewidth]{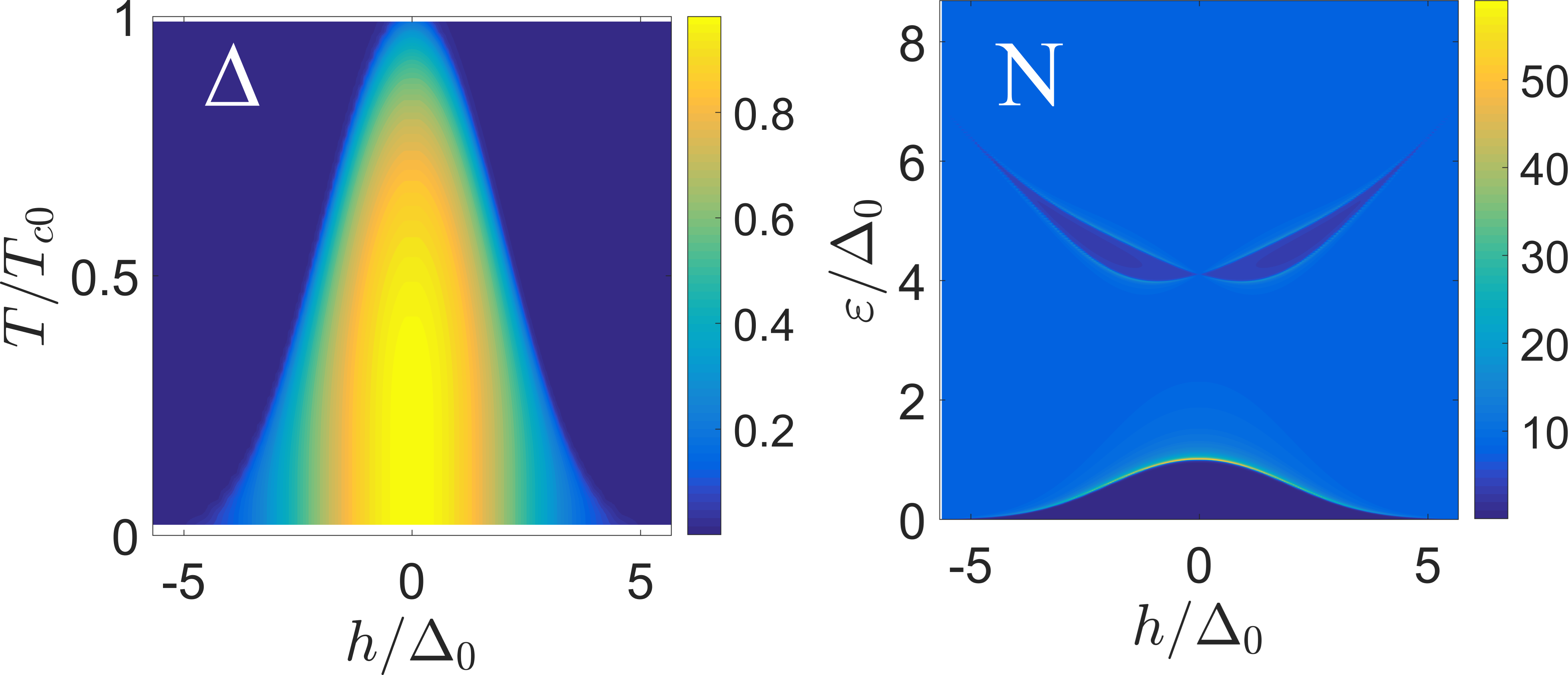}
  \put (-38,95) 
  {{\color{white} {\bf (b)}}}
   \put (-160,95) 
  {{\color{white} {\bf (a)}}}
  \end{array}$}
 \caption{\label{Fig:DOSOP}   
 (a) The order parameter $\Delta (T,h)/\Delta_0$.  
 (b) The density of states $N(\varepsilon,h)$ taken at $T=0.01$. In both panels $\beta=4\Delta_0$.   }
 \end{figure}   
 
 The equilibrium  GF $ \hat g_0 (\omega) e^{i\omega (t - t^\prime)}$ is determined by the stationary Eilenberger Eq.\ref{Eq:Eilenberger} in the imaginary time domain $t\to it $ with $\bm A=0$ and $\bm h = h \bm x = const$. 
 It has  an analytic solution \cite{supplement}. 
 With the equilibrium GF we can study the order parameter determined by the self-consistency equation 
 \begin{align}
 \Delta = \lambda \frac{\pi T}{4} \sum_{\omega} {\rm Tr} [\hat\tau_1  \hat g_0 (\omega) ]
 \end{align}          
 where $\omega$ runs over fermionic Matsubara frequencies. 
  The dependence  $\Delta(h,T)$  Fig.\ref{Fig:DOSOP}a for $\beta=4\Delta_0$ strongly exceeds the Chandrasekhar-Clogston limit \cite{Chandrasekhar1962,Clogston1962} $\Delta/\sqrt{2}$. Here we use the usual units of 
  $\Delta_0 =1.76 T_{c0}$, where $T_{c0}$ is the critical temperature at $h=0$. 
   The density of states is given by the analytic continuation $N(\varepsilon)= {\rm Re} {\rm Tr} [\hat \tau_3  \hat g_0 (-i\varepsilon) ]/4$. The function $N(\varepsilon,h)$ at $T=0.01 T_{c0}$ is shown in  Fig.\ref{Fig:DOSOP}b. It demonstrates the peaks both at the usual spectral gap and at the 
   "mirage" gap edges in accordance with the van Hove singularities of the spectrum (\ref{Eq:BogolubovSpecrum}).
   
\begin{figure}[htb!]
 \centerline{$
 \begin{array}{c} 
   \includegraphics[width=0.46\linewidth]
 {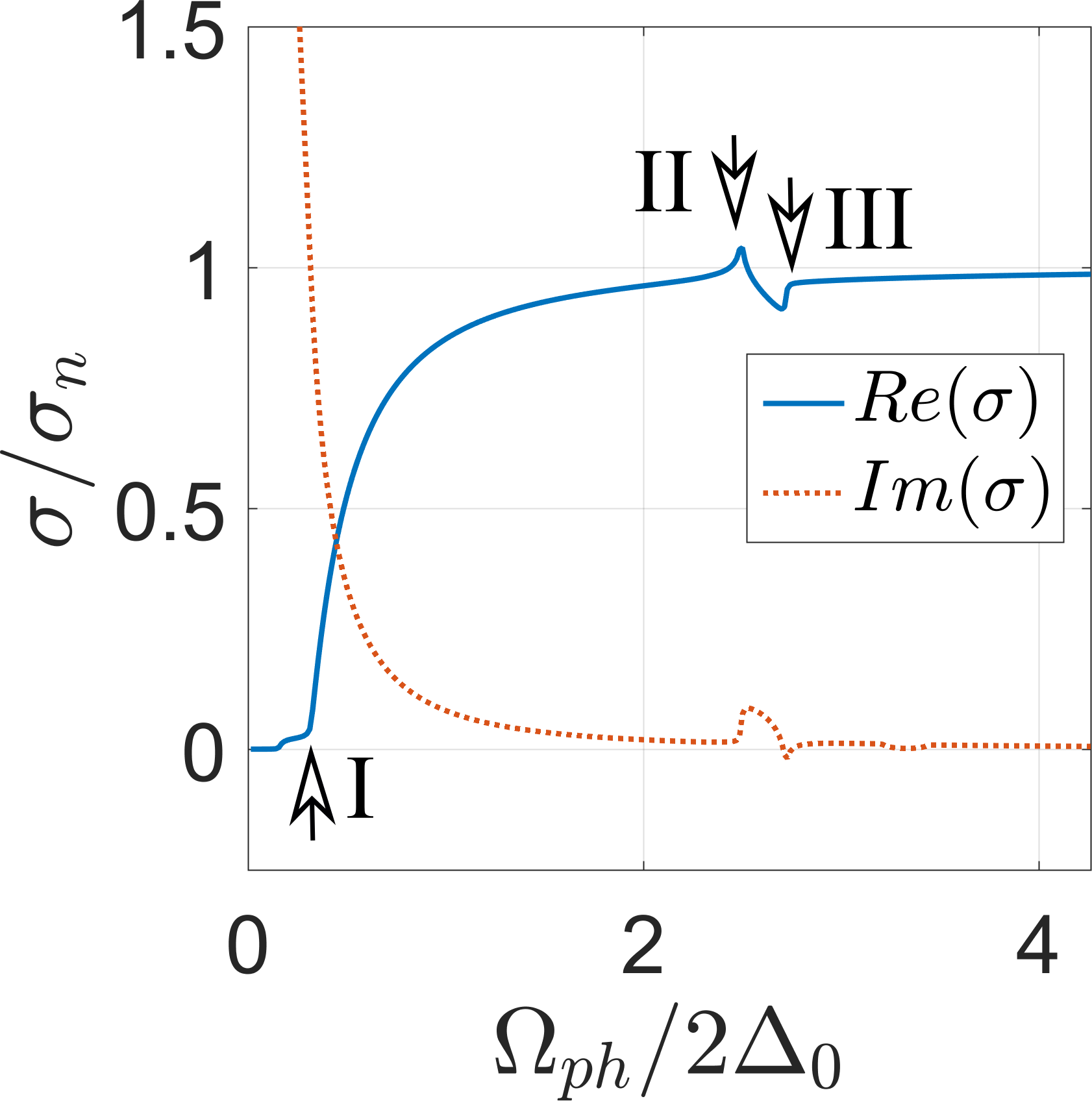}
 \;\;
 \includegraphics[width=0.52\linewidth]
 {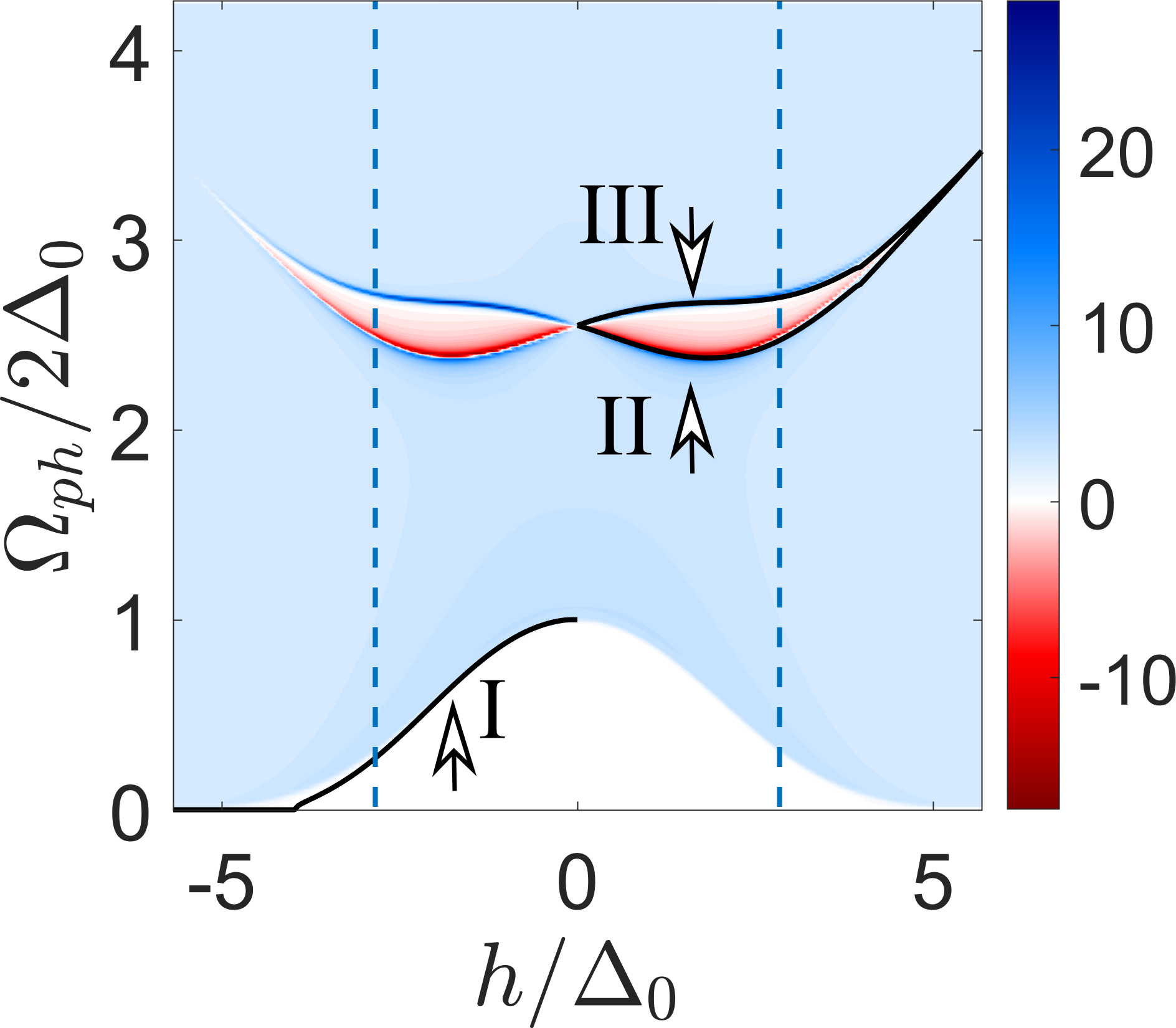}
  \put (-180,115) 
  {{\color{black} {\bf (a)}}}
 \put (-70,115) 
  {{\color{black} {\bf (b)}}}
 \\
  \includegraphics[width=0.46\linewidth]
  {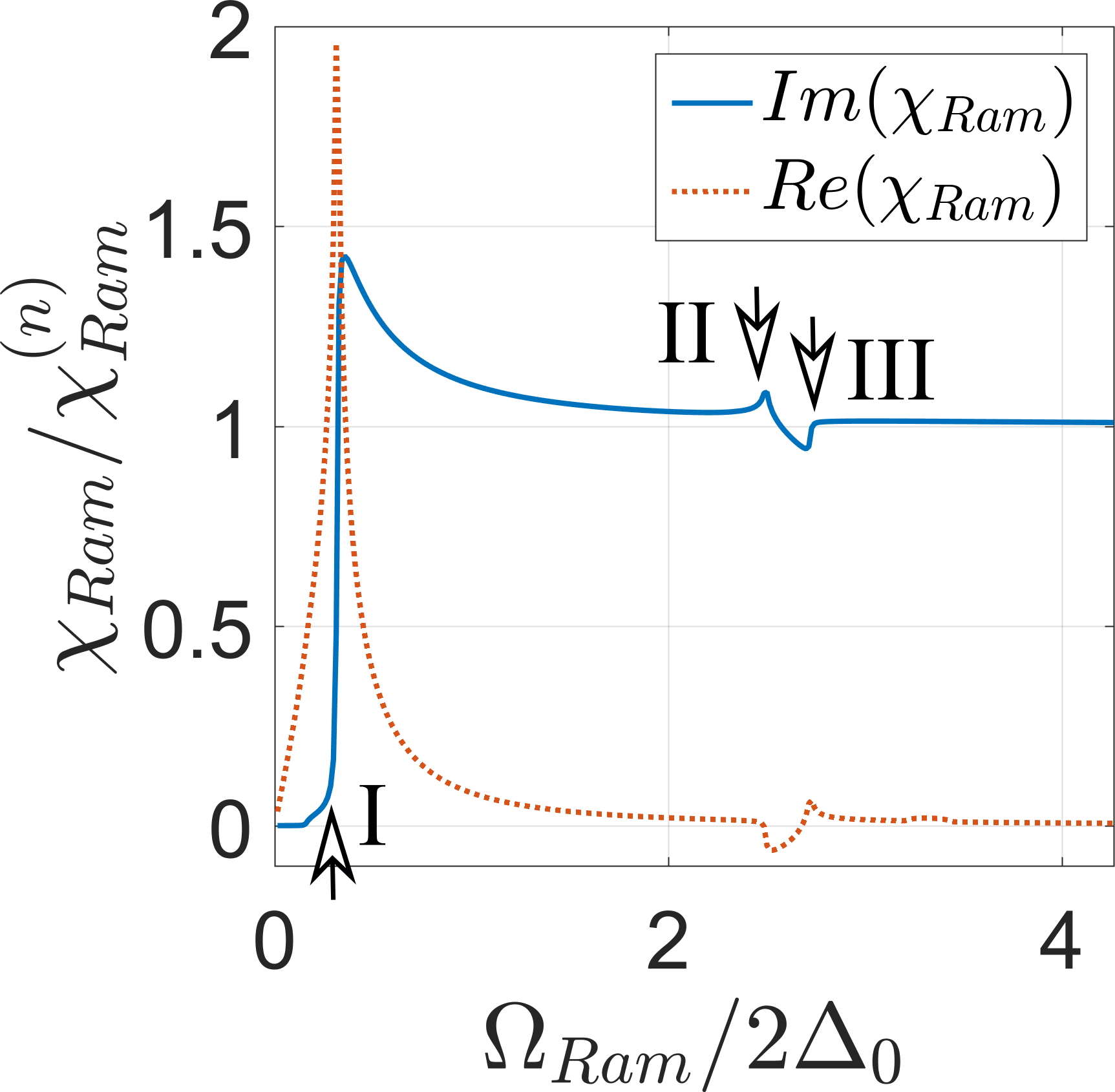}
  \;\;
   \includegraphics[width=0.52\linewidth]
  {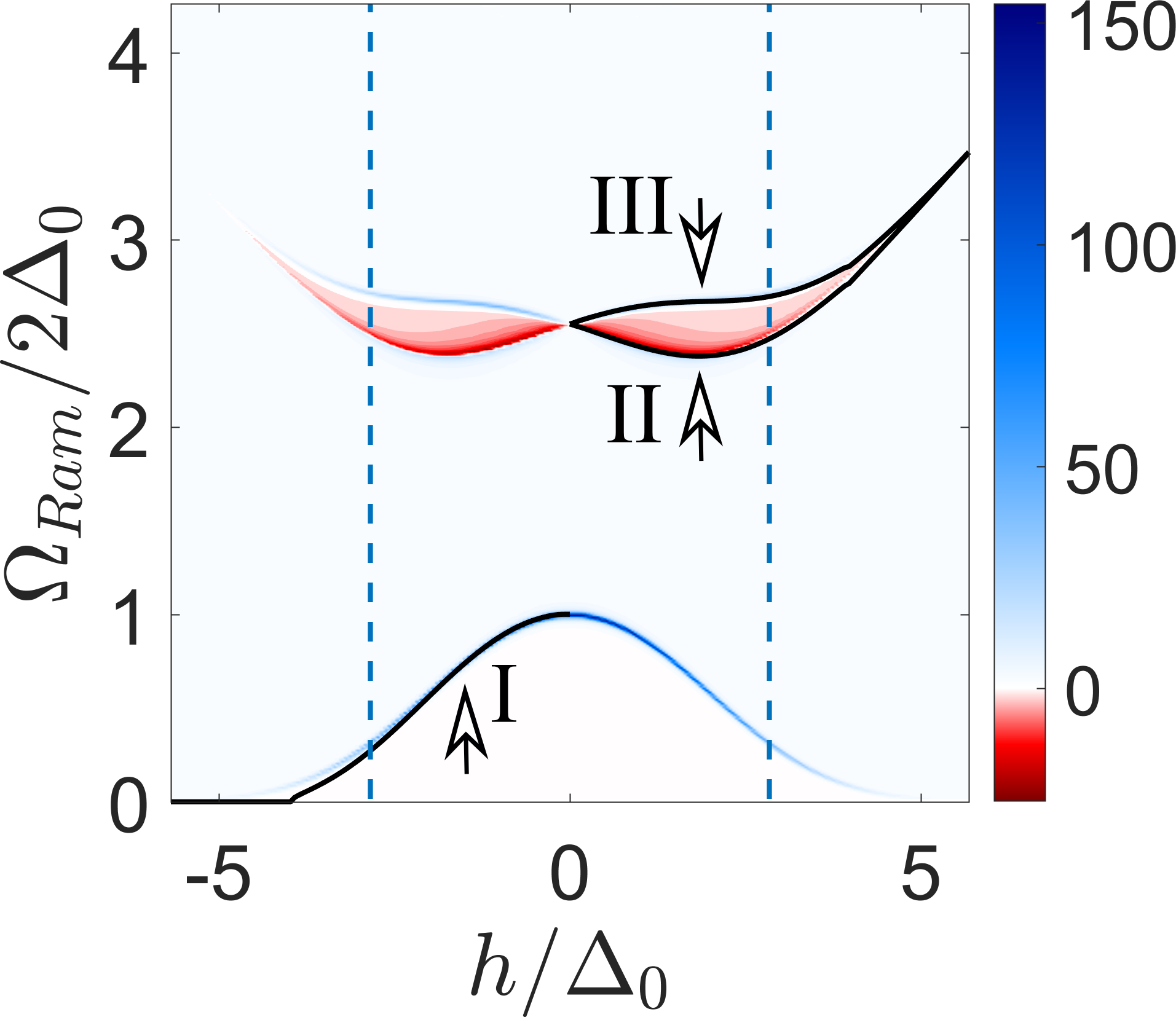}
 \put (-180,115) 
  {{\color{black} {\bf (c)}}}
 \put (-70,115) 
  {{\color{black} {\bf (d)}}}
 \end{array}$}
 \caption{\label{Fig:Conductivity} 
  (a) Conductivity $\sigma (\Omega_{ph})$ of the IS. (b) Frequency derivative of conductivity $d Re(\sigma)/d\Omega_{ph}$  .
 (c) Raman susceptibility  $\chi_{Ram}(\Omega_{Ram})$ of the IS. 
 (d) Frequency derivative of Raman susceptibility 
 ${\rm Im}( d\chi_{Ram}/d\Omega_{Ram})$. 
  Arrows $I$, $II$, $III$ point to the features corresponding to the resonant pair-breaking processes shown in Fig.\ref{Fig:Spectrum}a. 
 Black solid lines in (b,d) show the optical gaps $\Omega_{I}(h)$, $\Omega_{II}(h)$, $\Omega_{III}(h)$. In all panels $T=0.01 T_{c0}$, $\beta/\Delta_0=4$. In  (a,c) $h/\Delta_0=\pm 2.8$ which corresponds to the crossection along dashed lines in  (b,d). }
 \end{figure}

 { \it Linear response conductivity.}
  The conductivity tensor is given by the Fermi pocket average of the combination 
 $\sigma_{ij}=  \nu e^2 \langle v_{Fi} v_{Fj} \rangle_K \sigma $, where $\nu$ is the normal density of states, $e$ is the electron charge. The direction-independent conductivity amplitude $\sigma$  is determined by the  correction to the 
GF linear in the applied ac field $\bm A e^{i\Omega_{ph} t}$ 
which can be written in the form 
$ (\bm v_F \bm A) \hat g_A $. 
  Then the  conductivity is given by the expression in terms of the Keldysh component $\hat g_A^K$
 %
 \begin{align}
 \label{Eq:SigmaDirty}
 & \frac{\sigma }{\sigma_n} = 
 \frac{1}{16\Omega_{ph}} 
 \int_{-\infty}^{\infty} d\varepsilon {\rm Tr} [\tau_3\hat g_A^K]
 \end{align}
 where $\sigma_n=\tau_{imp}/4$ is the normal state
 conductivity. 
 The function $ \hat g_A $ can be found analytically   
 from the  Eilenberger Eq.\ref{Eq:Eilenberger}
linearized with respect to  $\bm A $. The particularly simple expression is obtained in the limit of large impurity scattering $\tau_{imp} \Delta \ll 1 $.
 We use this approximation to find $ \hat g_A $ for Matsubara frequencies and  and implementing the analytical continuation 
 to the real frequencies\cite{supplement}. 
 The resulting conductivity spectrum $\sigma (\Omega_{ph})$ is shown in the Fig.\ref{Fig:Conductivity}a.

{\it  Raman susceptibility}
 Now let us consider the Raman susceptibility 
  determined by the generalized  density-density correlation function\cite{devereaux2007inelastic,abrikosov1973theory,
klein1984theory,
falkovsky1993inelastic,
abrikosov1961raman} $ \nu\langle (\bm e_i \hat \gamma \bm e_f )^2 \rangle \chi_{Ram}  $. Here the prefactor is the Raman vertex determined by the initial $\bm A_i = e^{i\Omega_i t}A_i \bm e_i $ and final $\bm A_f = e^{i\Omega_f t} A_f \bm e_f $  photon polarizations. 
 The isotropic amplitude which we denote for brevity  $\chi_{Ram}$ can be found by solving the Eilenberger Eq.\ref{Eq:Eilenberger}  linearized with respect to the diamagnetic driving term $  (\bm A_i \hat \gamma \bm A_f^* ) e^{i\Omega_{Ram} t} $.
 The corresponding correction to the GF has the form 
 $   (\bm A_i \hat \gamma \bm A_f^* )  \hat g_{AA} $. It can be found analytically in the limit of large impurity scattering $\tau_{imp} \Delta \ll 1 $, see details in \cite{supplement}. Then the  Raman susceptibility is given by the expression in terms of the Keldysh component $\hat g_{AA}^K$
 \begin{align}
 \label{Eq:RamanDirty}
 \frac{\chi_{Ram}}{\chi_{Ram}^{(n)}} = 
 \frac{1}{16\Omega_{Ram}} 
 \int_{-\infty}^{\infty} d\varepsilon {\rm Tr} 
 [\hat g_{AA}^K]
 \end{align}
 where $\chi_{Ram}^{(n)} =\Omega_{Ram} \tau_{imp}/4 $ is the normal metal Raman susceptibility amplitude  in the dirty limit. Implementing the analytical continuation to the real frequencies\cite{supplement} $\Omega_{Ram} \to i \Omega_{Ram}$ yields the Raman susceptibility shown in the Fig. \ref{Fig:Conductivity}c,d.

Both $\sigma(\Omega_{ph})$ and $\chi_{Ram}(\Omega_{Ram})$ have three distinct features marked by the arrows $I$, $II$ and $III$ 
in Fig.\ref{Fig:Conductivity} corresponding to excitation processes across the optical gaps with the same numbering in Fig.\ref{Fig:Spectrum}a. To make these features  more visible we plot the derivatives 
$d {\rm Re}\sigma/d\Omega_{ph}$  in Fig.\ref{Fig:Conductivity}b and 
 $d \chi_{Ram}/d\Omega_{Ram}$  in Fig.\ref{Fig:Conductivity}d. One can see that the these derivatives have three peaks which perfectly coincide with the optical gaps shown by the black solid lines $\Omega_{I}(h)$, $\Omega_{II}(h)$, $\Omega_{III}(h)$. Note that these dependencies are symmetric by $h\to -h $. 

 {\it Superconducting laser with magnon pumping.} 
  The high-frequency electromagnetic pumping generates the population inversion in the spectrum of Bogolubov quasiparticles which  leads to the non-trivial effects such as the stimulated superconductivity 
 \cite{Eliashberg1970, Ivlev1971,PhysRevB.97.184516,klapwijk1977radiation,
klapwijk2020discovery}. However this type of the population inversion is not sufficient  for the superconducting laser, that is for the stimulated light mission due to the dipolar transitions between the different quasiparticle states. 
             
 \begin{figure}[htb!]
 \centerline{$
 \begin{array}{c} 
 \includegraphics[width=1.0\linewidth]
 {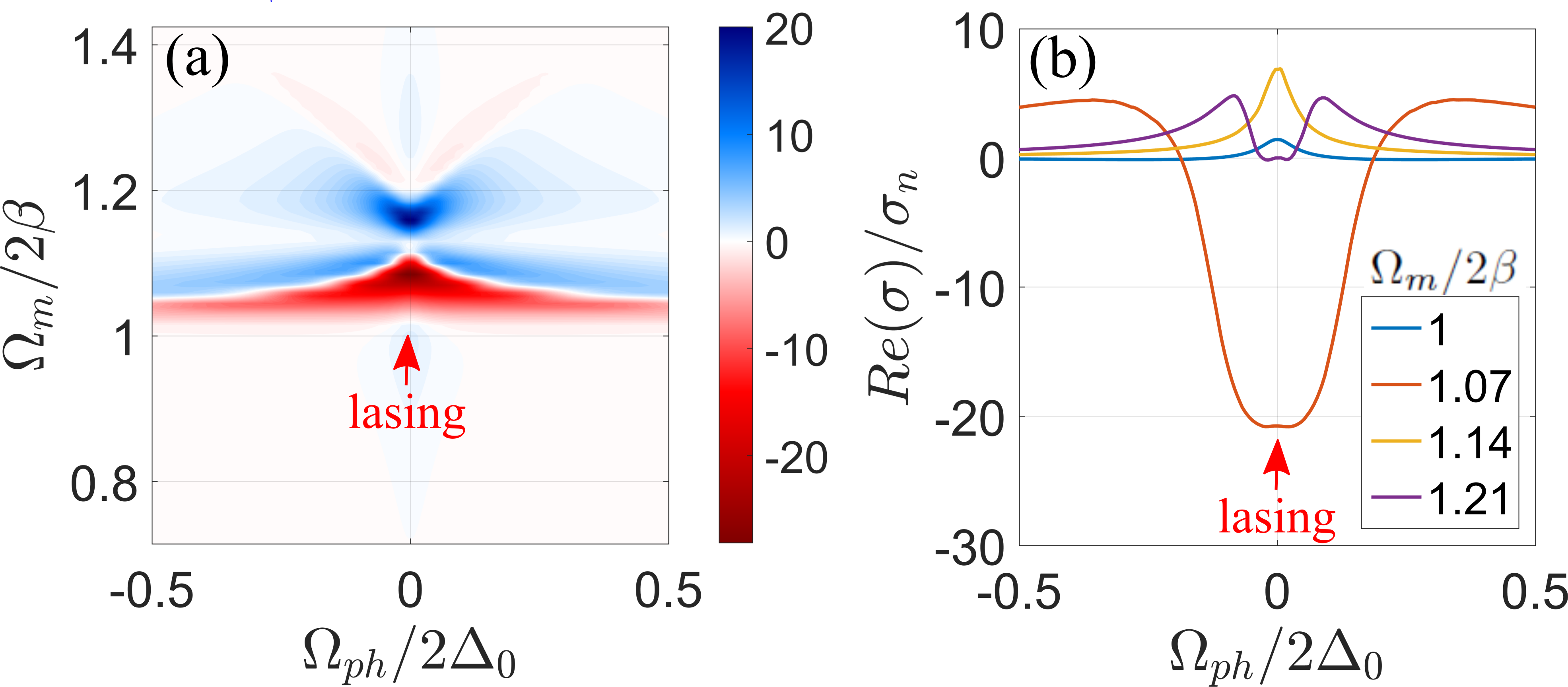} 
 \end{array}$}
 \caption{\label{Fig:SigmaStimulated} 
 $ {\rm Re} (\sigma) (\Omega_{ph},\Omega_{m})$ of the IS
 with magnon pumping 
 $\bm h = \bm x h_m\cos(\Omega_m t) $,   
 $h_m =  \sqrt{\Gamma T_{c0}} $, $\Gamma=0.01 T_{c0}$,  $T=0.01 T_{c0}$, $\beta=4\Delta_0$. The lasing regime corresponds to ${\rm Re} (\sigma) < 0$ in (a,b) according to the 
  mechanism in Fig.\ref{Fig:Spectrum}b. }
 \end{figure}   
The situation changes qualitatively in the presence of the Ising spin splitting in the Bogolubov spectrum shown in Fig.\ref{Fig:Spectrum}b. In this configuration one can generate a very strong quasiparticle imbalance by inducing transitions between spin-down and spin-up branches shown by the vertical dashed line corresponding to the magnonic pumping.  

 We assume that the superconductor is pumped by Zeeman field $\bm h=  h_m \cos(\Omega_m t) \bm x $ and calculate the  linear response conductivity $\sigma(\Omega_m,\Omega_{ph})$. 
 The total conductivity can be written as the two parts 
 $\sigma = \sigma_{eq} + \sigma_{hh}$. The first term  here $\sigma_{eq}$ is the  conductivity of the equilibrium superconductor  given by the Eq.\ref{Eq:SigmaDirty}. 
The second term $\sigma_{hh}$ is determined by the  correction to the GF linear in the applied ac field $\bm A e^{i\Omega_{ph} t}$ and of the second order in the oscillating Zeeman field amplitude
$ (\bm v_F \bm A)  h_m^2 \hat g_{Ahh} $. 
This correction  can be found analytically in the limit of large impurity scattering rate $\tau_{imp} \Delta \ll 1 $. 
 For this we use iterative solution\cite{supplement} of the 
 Eilenberger Eq.\ref{Eq:Eilenberger}   to get the stationary second-order correction driven by the Zeeman  $h_m^2 \hat g_{hh}$. We use at as a source to find the correction driven by the electromagnetic field $\hat g_{Ahh}$. 
Then, similarly to the usual conductivity (\ref{Eq:SigmaDirty}), the magnon-driven correction $\sigma_{hh}(\Omega_{ph})$ is given by
\begin{align}
 \label{Eq:SigmahhDirty}
 & \frac{\sigma_{hh} }{\sigma_n} = 
 \frac{h_m^2}{16\Omega_{ph}} 
 \int_{-\infty}^{\infty} d\varepsilon {\rm Tr} [\tau_3\hat g_{Ahh}^K]
 \end{align}
  Note that the magnitude of the correction $\sigma_{hh}$ is determined  by the ratio of pumping  strength and the rate of the inelastic relaxation $\Gamma$, so that $\sigma_{hh}/\sigma_n \sim h_m^2 \Gamma$. Since the inelastic relaxation is slow $\Gamma/ \Delta_0 \ll 1 $ we can achieve large   $\sigma_{hh} \sim \sigma_{eq}$ with weak pumping $h_m \sim \sqrt{\Gamma \Delta_0}$ .
  
 The total conductivity 
 ${\rm Re} \sigma (\Omega_{ph}, \Omega_m )$ is shown in the Fig.\ref{Fig:SigmaStimulated} for the narrow interval of magnon frequencies $\Omega_m\sim 2\beta$. There is a sharp negative peak at $\Omega_m \approx 2\beta$ and   $\Omega_{ph}\ll \Delta$ corresponding to the {\it superconducting laser} according to the mechanism shown in Fig.\ref{Fig:Spectrum}b. At larger pumping frequency $\Omega_m \approx 2.4\beta$ the hot spot in quasiparticle distribution becomes larger which leads to the stronger photon absorption than emission. In result there appears a sharp 
positive peak ${\rm Re}\sigma (\Omega_{ph}, \Omega_m) >0$. The peaks shown in Fig.\ref{Fig:SigmaStimulated}a at $\Omega_m \approx 2\beta$ are very sharp because of the  peculiar {\it resonant  nesting} of the branches \ref{Eq:BogolubovSpecrum}. Indeed  the interlevel distance is almost constant $min ( E_\uparrow + E_\downarrow) \approx 2\beta$ at $|\xi_k|<\beta$ which leads to the excitation of many momentum states by the same pumping frequency $\Omega_m \approx 2\beta$.    


 

 
To conclude, we predict  unusual optical properties of IS resulting from the interplay of the superconductivity, SOC and ferromagnetism. Multiple optical gaps in conductivity and Raman susceptibility offer the direct way to detect the finite-energy spin-triplet pairing correlations and the "mirage" gap\cite{tang2021controlling, tang2021magnetic} controlled by the in-plane magnetic field.  Moreover, we have shown that the Ising spin splitting of Bogolubov quasiparticle spectrum allows for the generation of quasiparticle population inversion driven by the oscillating Zeeman field.  This effect leads to the predicted superconducting laser which  potentially can be realized in VdW IS/FI structures.

Author thanks  Alexander Mel'nikov for useful discussions.   
This work was supported in part by the Russian science foundation.  
 \appendix
   
 \section{Calculation of the equilibrium Green function (GF)}
   
 The full GF is determined by the Gor'kov equation for imaginary frequeincies
 \begin{align} \label{AppEq:GFfull}
 & \hat G_0^{-1} \hat G =1 
 \\
 & \hat G_0^{-1} = \hat H + i\omega  
 \end{align}       
 
  We will need the quasiclassical GF which can be obtained from the full GF by integrating over the kinetic energy in the vicinity of the Fermi surface
 \begin{align}\label{Eq:QuasiclassicalGF}
 \hat g= \frac{i}{\pi} \int d\xi_k \hat\tau_3\hat G
 \end{align}
     
 The full GF determined by (\ref{AppEq:GFfull}) can be written as 
 \begin{align}
 \hat G = \frac{\hat C_0 + \xi_k \hat C_1 + 
 \xi_k^2 \hat C_2 + \xi_k^3 \hat C_3 }
 {(\xi_k^2 - E_1^2)(\xi_k^2 - E_2^{2} )}
 \end{align}
 where
 \begin{align}
 E_{1,2}= \sqrt{ h^2+\beta^2 -\Delta^2 -\omega^2 
 \pm 2i  \sqrt{\beta^2 (\Delta^2 + \omega^2) + h^2\omega^2} }   
 \end{align}
 Then the integration over $\xi_p$ yields
 \begin{align}
 & \int \frac{d\xi_k }{(\xi_k^2 - E_1^2)(\xi_k^2 - E_2^2)} = \frac{i \pi}{2E_1 E_2 (E_1 + E_2)}
 \\
 & \int d\xi_k \frac{\xi_k^2}{(\xi_k^2 - E_1^2)(\xi_k^2 - E_2^2)} =  - \frac{i\pi}{2(E_1 + E_2)}
 \end{align}
 Hence
 \begin{align}
 \hat g_0 = \frac{1}{2(E_1+E_2)}\left(\hat C_2 - \frac{\hat C_0 }{E_1E_2}\right)
 \end{align}
        
It is instructive to check the structure of quasiclassical GF in spin-Nambu space. In general it is given by 
 \begin{align}
 & \hat g_0 = 
 g_{z0}  \hat\sigma_z\hat\tau_0 + 
 g_{y2} \hat\sigma_y\hat\tau_2 +
 \\ \nonumber
 &  (g_{01} \hat\sigma_0 + g_{x1} \hat\sigma_x)\hat\tau_1 + (g_{03} \hat\sigma_0 + g_{x3} \hat\sigma_x)\hat\tau_3
 \end{align}
The last four terms in the r.h.s. are the usual ones for the superconductor with the Zeeman field. The first two terms appear due to the interplay of the Ising SOC and the Zeeman spin splitting fields. 
        


\section{Calculation of the conductivity and Raman susceptibility}
    
 To find the conductivity 
 %
 we need to determine the 
 linear correction to the GF 
 $ (\bm v_F \bm A) \hat g_A( \omega_+,\omega) $ generated by the applied ac field $\bm A e^{i\Omega_{ph} t}$ where $\omega_+= \omega + \Omega_{ph}$. 
For that we need to solve the linearized Eilenberger equation for imaginary frequencies
 \begin{align} \label{Eq:EilenbergerLinA_app}
 & i [ \hat\tau_3 \hat g_0 (\omega) -  \hat g_0 (\omega_+) \hat\tau_3 ]  = 
 \\ \nonumber
 & [\hat\Lambda , \hat g_A]
  + [ \omega_+\hat\tau_3  \hat g_A  -  \omega\hat g_A  \hat\tau_3 ] +
  \\ \nonumber
  &  \tau_{imp}^{-1} [ \hat g_0 (\omega_+) 
  \hat g_A  - \hat g_A \hat g_0 (\omega) ]
 \end{align} 
 The solution can be found analytically in the limit of $\tau_{imp} \Delta_0 \ll 1$ in the form  
 \begin{align} \label{Eq:gA_app}
 \hat g_A = \frac{i \tau_{imp}}{2}
 [\hat g(\omega_+) \hat\tau_3 \hat g(\omega) - \hat\sigma_0\hat\tau_3]
 \end{align}
 Implementing the analytical continuation to the real frequencies we obtain the Keldysh component $\hat g_A^K$ and hence the conductivity.

 
 The Raman susceptibility 
 is given by the linear response relation written in terms of the quasiclassical GF correction
 linear in the diamagnetic term in the Eilenberger equation  $ (\bm A_i \hat \gamma \bm A_f^* )  \hat g_{AA} (\omega_+,\omega)$
 where $\omega_+ = \omega + \Omega_{Ram}$. 
 The Eilenberger equation for the correction $\hat g_{AA}$ reads as 
 \begin{align} \label{Eq:EilenbergerRaman}
 &  \hat g_0(\omega) - \hat g_0(\omega_+) 
 = 
 \\ \nonumber
 & [\hat\Lambda , \hat g_{AA}]
 +
 [ \omega_+\hat\tau_3  \hat g_{AA}  -  \omega\hat g_{AA}  \hat\tau_3 ] +
 \\ \nonumber
 & \tau_{imp}^{-1} [ \hat g_0 (\omega_+) \hat g_{AA} -  \hat g_{AA} \hat g_0 (\omega) ]
 \end{align}  
 where we take into account that $\langle (\bm e_i \hat \gamma \bm e_f )  \rangle=0$. 
  %
 The solution can be found analytically in the limit of $\tau_{imp} \Delta_0 \ll 1$ in the form  
 \begin{align} \label{Eq:gAA_app}
 \hat g_{AA} = \frac{i \tau_{imp}}{2}
 [\hat g(\omega_+) \hat g(\omega) - \hat\sigma_0\hat\tau_0]
 \end{align}
 Implementing the analytical continuation to the real frequencies we obtain the Keldysh component $\hat g_{AA}^K$ and hence the Raman susceptibility.  
  
 \section{Conductivity of the Ising superconductor with magnon pumping}
 \label{Sec:MagnonPumpingCalculation}
  
  We assume that the Ising superconductor is subjected to the ac electromagnetic field $\bm A e^{i\Omega_{ph} t}$ and the Zeeman field with two tones 
  $h(t) = h_m (e^{i\Omega_{m1} t} + e^{i\Omega_{m2} t})$. In order to construct correctly the analytical to real frequencies we need to consider from the beginning the general case with $\Omega_{m1}\neq - \Omega_{m2}$. 
  We introduce the following notation for frequencies 
     \begin{align}
  & \omega_1= \omega,
  \\ \nonumber
  & \omega_2= \omega + \Omega_{m1},
   \\ \nonumber
  & \omega_3= \omega + \Omega_{m2}
  \\ \nonumber
  & \omega_4= \omega + \Omega_{m1} + \Omega_{m2} 
  \\ \nonumber
  & \omega_5=  \omega + \Omega_{ph}
   \\ \nonumber
  & \omega_6= \omega + \Omega_{m1} + \Omega_{m2} + \Omega_{ph}.
  \end{align}
  First, we find the second-order correction to the GF $\hat g_{hh} (41) \sim h_m^2 $, where we introduce notation $\hat g_{hh} (41) \equiv \hat g_{hh} (\omega_4,\omega_1)$, so that the in the time domain it is $\hat g_{hh} (t,t^\prime) =  \hat g_{hh} (\omega_4,\omega_1) e^{i( \omega_4 t - \omega_1 t^\prime ) }$. 
   After the analytical continuation we will put 
 $\Omega_{m1} = - \Omega_{m2}= \Omega_m$ to get the stationary correction $\hat g_{hh} (t,t^\prime) =
 \hat g_{hh} (t-t^\prime)$.  
 %
 We search this correction by solving the Usadel equation by iterations with respect to the Zeeman field in the imaginary frequencies.
   We start with the first-order correction 
  $\hat g_h = h_{m} (\hat g_x \hat \sigma_x + \hat g_y \hat \sigma_x)$. 
  For that we need to solve the linearized Eilenberger equation for imaginary frequencies
 \begin{align} \label{Eq:EilenbergerLinh_app}
 & i \hat\sigma_x[ \hat\tau_3 \hat g_0 (\omega) -  \hat g_0 (\omega_+) \hat\tau_3 ]  = 
 \\ \nonumber
 & [\hat\Lambda , \hat g_h]
  + [ \omega_+\hat\tau_3  \hat g_h - \omega\hat g_h  \hat\tau_3 ] 
 \end{align} 
  where $\omega_+ = \omega_2$ for $\hat g_h (21)$ or $\omega_+ = \omega_3$ for $\hat g_h (31)$.     
  The solution is given by 
 \begin{align}
 & \hat g_x (21)= \frac{i(s_1 + s_2)}{ (s_1+s_2)^2 + 4\beta^2} [ \hat g_0 (2) \hat\tau_3 \hat g_0 (1) - \hat\tau_3 ] 
 \\
 & \hat g_y (21)= \frac{2i\beta}{ (s_1+s_2)^2 + 4\beta^2} 
 [ \hat\tau_3 \hat g_0 (1) - \hat g_0 (2)\hat\tau_3 ]
  \end{align}
 For $\hat g_h (31)$ we need to replace $2\to 3$. 
 The second order correction $\hat g_{hh} (41)$ can be found by solving the equation
 \begin{align}
 & s_4 \hat g_0(4) \hat g_{hh} -
  s_1 \hat g_{hh}  \hat g_0(1)
  = 
  \\ \nonumber
  & i [\hat\tau_3 \hat g_x (21) - \hat g_x (43) \hat\tau_3]  
  + 
  i [\hat\tau_3 \hat g_x (31) - \hat g_x (42) \hat\tau_3]
 \end{align}
   It is supplemented by the normalization condition 
   \begin{align}
    & \hat g_0(4)\hat g_{hh} + \hat g_{hh}\hat g_0(1) + \\ \nonumber
    & \hat g_x(43)\hat g_x(31) +
    \hat g_x(42)\hat g_x(21)
    +
    \\ \nonumber
    & \hat g_y(43)\hat g_y(31) +
    \hat g_y(42)\hat g_y(21)
    =0
   \end{align}
   The solution is given by 
  \begin{align}
 & (s_1 + s_4)\hat g_{hh} (41)= 
  \\ \nonumber
  &   i  \hat g_0(4) 
  [ \hat\tau_3 ( \hat g_x (21) +  \hat g_x (31)) - 
  (\hat g_x (43) + \hat g_x (42)) \hat\tau_3 ]   
  \\ \nonumber
  & - s_1 \hat g_0(4) 
  [ \hat g_x (42)\hat g_x (21) + \hat g_x (43)\hat g_x (31)] 
\\ \nonumber
  & - s_1 \hat g_0(4) 
  [ \hat g_y (42)\hat g_y (21) + \hat g_y (43)\hat g_y (31)] 
  \end{align}

 {
 To find the contribution to electric current in the presence of magnon pumping we need to find the  anisotropic correction to the GF which is linear in $\bm A$ and of the second order in Zeeman field
 %
 $ (\bm v_F \bm A) h_m^2 \hat g_{Ahh}$.
 We consider the limit of strong impurity scattering when the anisotropic correction driven by the field $\bm A e^{i\Omega_{ph}t}$ to the 
 GF is in general given by $(\bm v_F \bm A) \hat g_A$
 \begin{align} \label{Eq:gAgeneral_App}
 & \hat g_A (t,t^\prime)=  
 \\ \nonumber
 & \frac{i  \tau_{imp}}{2}
[ \hat g (t,t_1) e^{i\Omega_{ph}t_1} \hat\tau_3  \circ \hat g(t_1,t^\prime) - \hat\tau_3\hat\sigma_0 \delta(t-t^\prime) e^{i\Omega_{ph}t}] 
 \end{align}
 where $\hat g(t_1,t^\prime)$ is in general the non-equilibrium the GF in the absence of electromagnetic field. 
 Taking into account that in the presence of the magnon pumping $\hat g = \hat g_0 + \hat g_{hh}$ and going to the Fourier representation we get from (\ref{Eq:gAgeneral_App})  
 $\hat g_{Ahh}(t,t^\prime)= \hat g_{Ahh} (61)  e^{i(\omega_6 t-\omega_1 t^\prime)}$ where
 \begin{align} \label{Eq:gAhh_app}
 \hat g_{Ahh} (61)= 
  \frac{i \tau_{imp}}{2}
 [\hat g_0(6) 
 \tau_3 \hat g_{hh}(41) 
 + 
 \hat g_{hh}(65) \tau_3  \hat g_0(1)]
 \end{align}
 Using this expression we can use the analytical continuation to real energies to find the Keldysh component $\hat g_{Ahh}^K$ which determines the
 correction to the conductivity $\sigma_{hh}$. 
  }
 
 %


   \section{Analytical continuation} 
 \label{Sec:AnalyticalContinuation}
 In order to find the real-frequency response we need to implement the analytical continuation. 
  The analytical continuation of the sum by Matsubara frequencies is determined according to the general rule \cite{kopnin2001theory}
 \begin{align} \label{Eq:AnalyticalContinuationGen}
 & T\sum_\omega \hat g_1(\omega_1)\hat g_2(\omega_2)
 \to  
 \\ \nonumber
 &  \int \frac{d\varepsilon}{4\pi i}  
 n_0(\varepsilon_1)
 \left[ \hat g_1( -i \varepsilon^R_1)
 - 
 \hat g_1(-i \varepsilon^A_1)
 \right] \hat g_2(-i \varepsilon^A_2) 
 +
 \\ \nonumber 
 & \int \frac{d\varepsilon}{4\pi i}  
 n_0(\varepsilon_2)
 \hat g_1(-i \varepsilon^R_1) 
 \left[ \hat g_2( -i \varepsilon^R_2)
 - 
 \hat g_2(-i \varepsilon^A_2)
 \right] 
  \end{align}
 where $n_0(\varepsilon) = \tanh (\varepsilon/2T ) $ is the equilibrium  
 distribution function. In the r.h.s. of (\ref{Eq:AnalyticalContinuationGen}) we substitute 
  $\varepsilon_1 = \varepsilon $, $\varepsilon_2 = \varepsilon + \Omega$ 
  and  $\varepsilon^{R/A} = \varepsilon \pm  i\Gamma$. Here the term with $\Gamma>0$ is added to   
  shift of the integration  contour into the corresponding half-plane.
   
   \bibliography{refsHiggs}

 \end{document}